\documentclass[journal,fleqn,10pt,twocolumn,twoside]{IEEEtran}

\usepackage{amsmath,amssymb}
\usepackage{graphicx}
\usepackage{textcomp,gensymb}                 
\usepackage{multirow}                         
\usepackage{cite}                             
\usepackage[linesnumbered,ruled]{algorithm2e} 
\SetKwInOut{Input}{Input}
\SetKwInOut{Output}{Output}
\usepackage{comment}                          
\usepackage{xcolor}                           

\usepackage{tikz}
\definecolor{lime}{HTML}{A6CE39}
\DeclareRobustCommand{\orcidicon}{%
	\begin{tikzpicture}
	\draw[lime, fill=lime] (0,0) 
	circle [radius=0.16] 
	node[white] {{\fontfamily{qag}\selectfont \tiny ID}};
	\draw[white, fill=white] (-0.0625,0.095) 
	circle [radius=0.007];
	\end{tikzpicture}
	\hspace{-2mm}
}
\foreach \x in {A, ..., Z}{%
	\expandafter\xdef\csname orcid\x\endcsname{\noexpand\href{https://orcid.org/\csname orcidauthor\x\endcsname}{\noexpand\orcidicon}}
}

\usepackage{hyperref} 

\usepackage{tikz}
\usetikzlibrary{trees}
\usepackage[flushleft]{threeparttable} 
\usepackage[pscoord]{eso-pic}
\definecolor{ao}{rgb}{0.0, 0.0, 1.0}
\IEEEoverridecommandlockouts

\begin{document}

\title{Channel Modeling for 60 GHz Fixed mmWave O2I and O2O Uplink with Angular Misalignment}

\author{Nitisha~Singh\orcidA{},
Sahaj~K.~Jha\orcidB{},
Aniruddha~Chandra\orcidC{}, 
Radek~Zavorka,
Petr Horky,
Tomas Mikulasek\orcidD{},\\
Jiri Blumenstein\orcidX{},
Ales~Prokes\orcidE{},
Jaroslaw~Wojtun\orcidF{},
Jan~M.~Kelner\orcidG{}, 
Cezary~Ziolkowski\orcidH{}
%
%
%
\thanks{This work was developed within a framework of the research grants: project no. 23-04304L sponsored by the Czech Science Foundation, MubaMilWave no. 2021/43/I/ST7/03294 funded by National Science Centre, Poland under the OPUS call in the Weave programme, grant no. UGB/22-748/2024/WAT sponsored by the Military University of Technology, and chips-to-startup (C2S) program no. EE-9/2/2021-R\&D-E sponsored by MeitY, Government of India.
}

\thanks{N. Singh, S. K. Jha and A. Chandra are with the Department of Electronics and Communication Engineering, National Institute of Technology Durgapur, West Bengal-713209, India (e-mail: aniruddha.chandra@ieee.org).}
\thanks{R. Zavorka, P. Horky, T. Mikulasek, J. Blumenstein and A. Prokes are with the Department of Radio Electronics, Brno University of Technology, 61600 Brno, Czech Republic.}
\thanks{J. Wojtun, J. M. Kelner and C. Ziolkowski are with the Institute of Communications Systems, Faculty of Electronics, Military University of Technology, Warsaw, Poland.}
}

\maketitle
\markboth{IEEE Antennas and Wireless Propagation Letters, Vol. 23, No. 5, pp. 1653-1657, May 2024, \url{https://doi.org/10.1109/LAWP.2024.3365568}}{SINGH \MakeLowercase{\textit{et al.}}: 60 GHz MMWAVE CHANNEL MODEL WITH MISALIGNMENT IN O2I AND O2O SCENARIOS}

\begin{abstract}
In this paper, we examine the effect of misalignment angle on cluster-based power delay profile modeling for a 60 GHz millimeter-wave (mmWave) uplink. The analysis uses real-world data, where fixed uplink scenarios are realized by placing the transmitter at ground level and the receiver at the building level. Both outdoor-to-indoor (O2I) and outdoor-to-outdoor (O2O) scenarios are studied. Using the misalignment angle and the scenario as inputs, we propose a statistical power delay profile (PDP) simulation algorithm based on the Saleh-Valenzuela (SV) model. Different goodness-of-fit metrics reveal that our proposed algorithm is robust to both O2I and O2O scenarios and can approximate the PDPs fairly well, even in case of misalignment.  
\end{abstract}


\section{Introduction}
Realizing more than $20$ Gbps data rates for sixth generation ($6$G) dream use cases, such as holographic communication or tactile internet, is not possible without climbing the spectrum ladder till the millimeter wave (mmWave) frequencies \cite{dang2020should}. The $3$rd generation partnership project ($3$GPP) is currently creating provisions in the fifth generation new radio ($5$G NR) frequency range $2$ (FR$2$) up to $71$ GHz with a special priority given to $60$ GHz \cite{3gppIntel}. This unlicensed band alleviates hefty spectrum fees, offers a contiguous band avoiding complex carrier aggregation, and provides an order of higher bandwidth than sub-$6$ GHz FR$1$.

The $60$ GHz band suffers from regular mmWave disadvantages, i.e., high directionality and large attenuation, limiting $60$ GHz propagation investigations concentrated primarily on indoor and short-range communication, as evident from some articles \cite{awpl1,awpl2} that appeared in the back issues of the current journal. In most outdoor $60$ GHz sounding studies for link lengths of $\sim$100m \cite{aslam2020,kim2021,du2022}, measurement results were compared against the standard 3GPP urban micro (UMi) street canyon model, except \cite{de2022}, where a physics-based model was developed. On the other hand, geometry-based stochastic modeling was attempted in \cite{kelner2021}, while authors in \cite{lubke2021,jun2022} focused on validating ray tracing based simulation against $60$ GHz outdoor measurements.  

This motivated us to test the suitability of the band for outdoor low-mobility links between an user equipment (UE) and an access point (AP) at a customer premise; often placed on a rooftop, a window, or mounted on a wall inside a building \cite{GSMAmmW}. The cellular uplink coverage footprint is smaller than the downlink coverage footprint, irrespective of the center frequency. This is because the AP antenna gain is higher than the one in UE. The presence of antenna arrays at AP increases the gain difference as well as the coverage gap for $60$ GHz mmWave. Thus, it is important to model $60$ GHz uplink, which defines the lower bound of a cell coverage radius. Further, narrow beamwidth affects the angular property of the uplink, and it is worth studying how misalignment can impact the received signal \cite{10}.

In this paper, we investigate the misalignment of 60 GHz outdoor uplinks in outdoor-to-indoor (O2I) and outdoor-to-outdoor (O2O) scenarios. We present an algorithm based on the Saleh Valenzuela (S-V) model, which takes the O2I or O2O scenario and misalignment range as inputs and simulates a power delay profile (PDP). Moreover, we employ statistical metrics such as the root-mean-square (RMS) delay spread, correlation, and the two sample Kolmogorov-Smirnov (K-S) test to show that the generated PDPs closely follow the PDPs in a given case. The major contributions of this paper include:
\begin{itemize}
    \item We show, based on actual field test data, PDP for both the outdoor scenarios (O2I/ O2O) can be modeled through an unified approach. It may be noted that separate measurement based channel models for O2I and O2O cases exist in the current literature \cite[Table 1]{kohli2022outdoor}. Also, we focus on rms delay spread, which is a dense multipath component (MPC) dominated parameter. Although MPCs outnumber specular components (SCs) in indoor mmWave propagation, the contribution of MPCs is non-negligible for outdoor propagation environments. 
    
    \item We show that S-V model, with suitable modification, can be applied to characterize outdoor long-distance 60 GHz mmWave link. As far as this specific band is concerned, S-V model implementation was restricted to short distances, e.g., a 60 GHz desktop channel \cite{liu2007characterization}.
    
    \item We were able to show that power angular spectrum (PAS) can be broadly divided into two angular regions, namely (0$^\circ$, 10$^\circ$] and (10$^\circ$, 25$^\circ$], over which the PDP parameters remain fairly constant, which greatly simplifies angle-dependent channel model over the sophisticated angular clustering algorithms  \cite{lyu2021angular}.
\end{itemize}

\begin{figure}[t!]
    \centering
    \includegraphics[width=3in,clip,keepaspectratio]{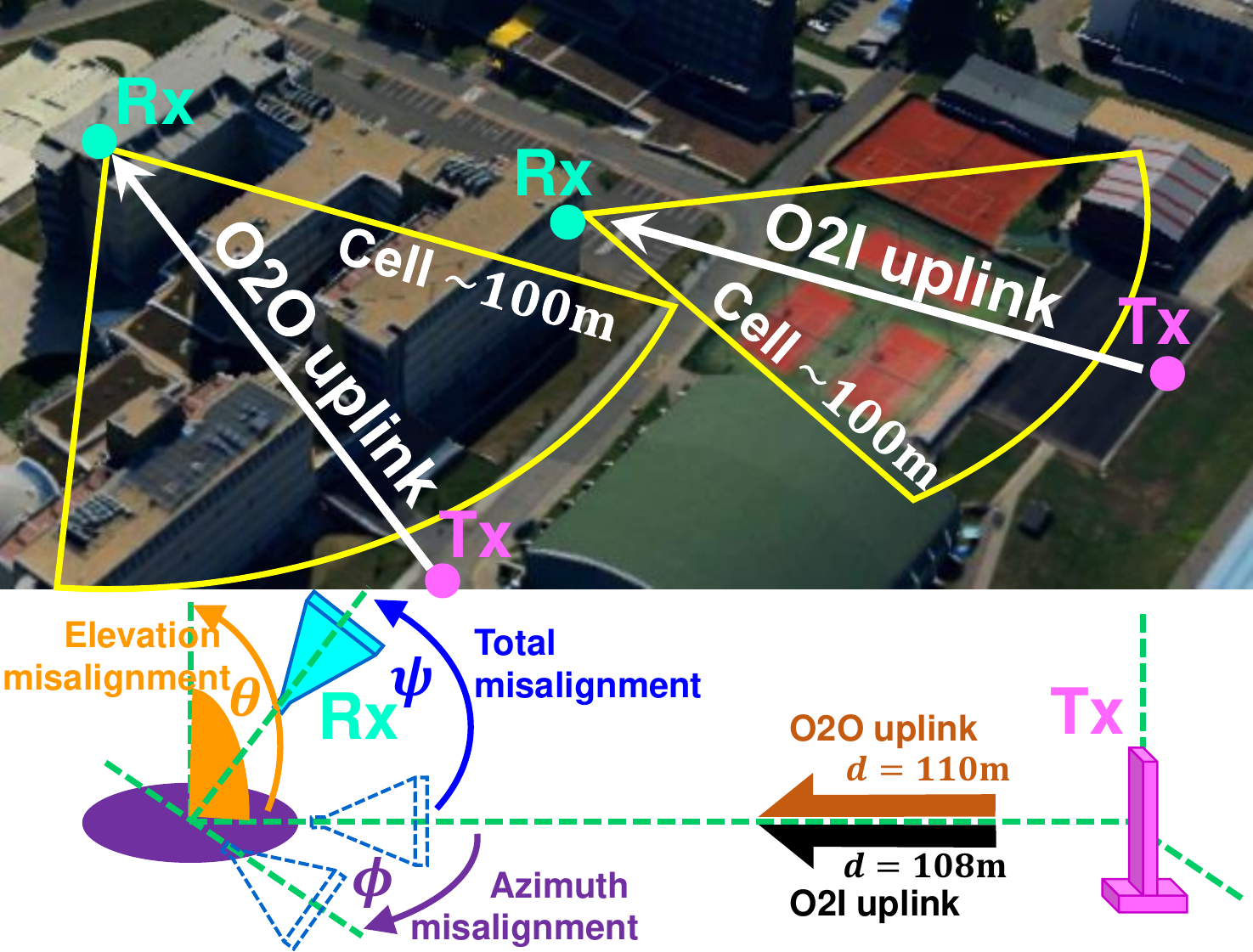}
    \caption{Details of field measurement: [top] Aerial view of the test site showing the position of Rx in FEKT building (rooftop for O2O/ wall-mounted for O2I) and position of Tx at cell edge ($\approx 100$m) at ground level, [bottom] Total Rx misalignment angle calculation from azimuth and elevation misalignment. Aerial view image courtesy: mapy.cz.}
    \label{fig:Fig01}
\end{figure}

\section{Field Measurement and Post-Processing}
Fig. \ref{fig:Fig01} describes the O2I and O2O environments for field measurements conducted at the campus of Brno University of Technology, Brno, Czech Republic (49$^\circ$13$'$37$''$N, 16$^\circ$34$'$28$''$E). For emulating an uplink scenario, the transmitter (Tx) was placed at ground level at the edge of a fictitious cell of radius $\approx$ 100m, typical for mmWave urban micro implementations. The receiver (Rx) was placed at the building level, inside a window for the O2I case, and on the rooftop for the O2O case. On the Tx side, we used an analog signal generator (model: Agilent E8257D) and on the Rx side, a scalar signal analyzer (model: Rohde \& Schwarz FSUP50) was used like our earlier measurement campaigns \cite{ChandraNTMS,RahmanVNC,ChandraACCESS,ShuklaJWCN}, thus, only the magnitude data is recorded \cite{beammisalignment}. The power received was recorded for angular combinations over a frequency range of 56 GHz to 64 GHz with a frequency resolution of 0.1 GHz and a minimum angular resolution of 5$^\circ$. While the Tx was realized with a custom-built one-sided SIW slot antenna \cite{mikulasek2015siw}, the Rx was a standard gain directional horn antenna (model: Quinstar QWH-VPRR00 50–75 GHz). The Rx antenna was fitted on an astrotracker motorized mount for angular sweep control. The half-power beamwidth of the antenna used is 10$^\circ$, which is larger than the angular step variation and ensures sufficient angular space to receive distinguished MPCs.

For the O2I scenario, the Tx was positioned at a tennis court (1.6 m above ground) and the Rx was stationed inside Technicka 12 building (13.5 m above ground level). The receiver is rotated horizontally (-25$^\circ$ to 35$^\circ$) and vertically from (-5$^\circ$ to 5$^\circ$) 

For the O2O scenario, both the Tx and Rx are placed outside. The height of the Tx is 1.6 m. The Rx is positioned at the top of Technicka 12 building (18 m above the ground). The Rx is rotated horizontally (-22.5$^\circ$ to 22.5$^\circ$) for a vertical misalignment of 4.33$^\circ$ and -4.33$^\circ$, and from (-25$^\circ$ to 25$^\circ$) for vertical misalignment of 8.66$^\circ$ and -8.66$^\circ$.  

%
%

\section{Channel Model Development}
\begin{figure}[ht]
    \centering
    \includegraphics[width=3.5in,clip,keepaspectratio]{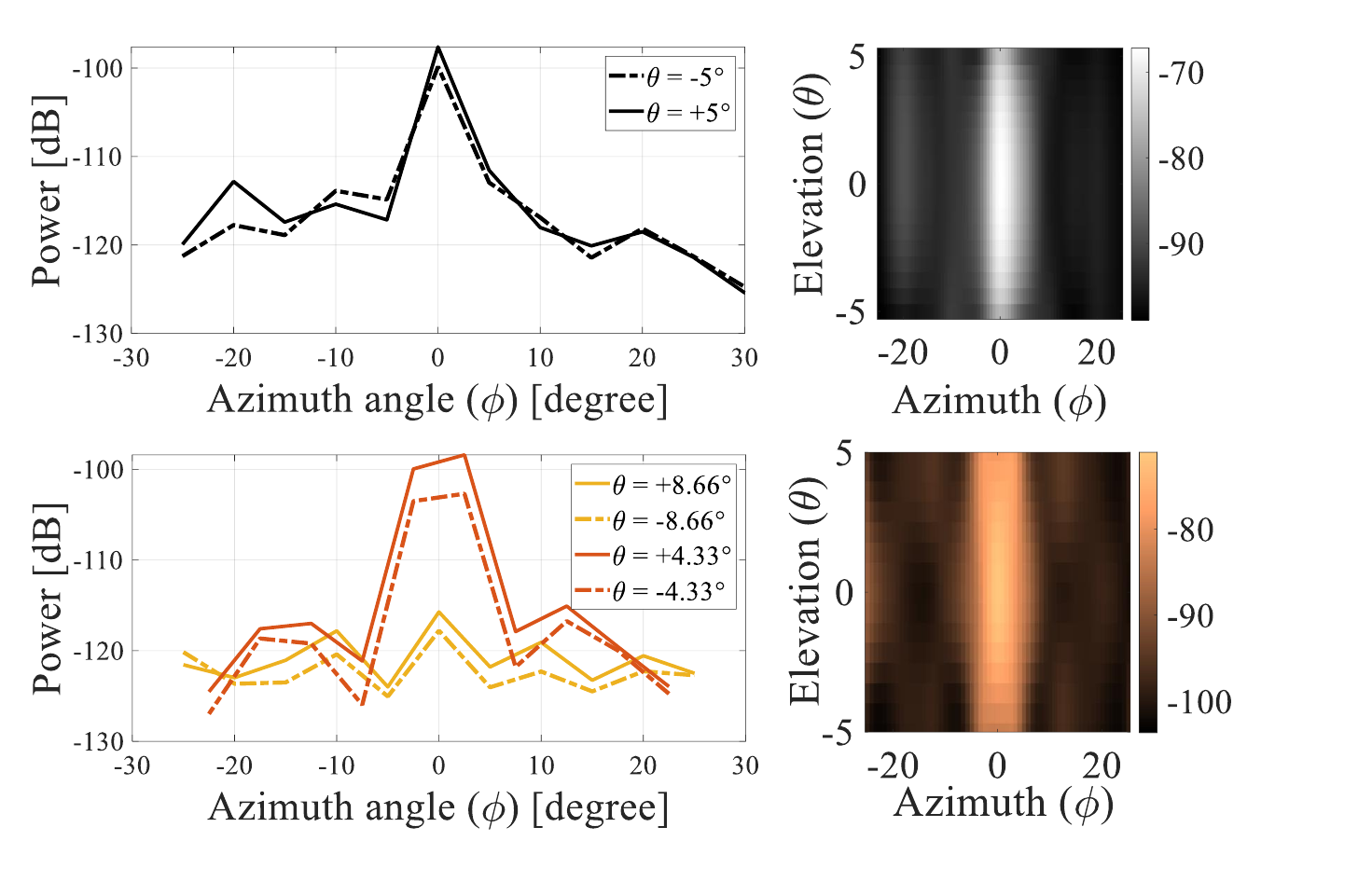}
    \caption{PAP$(\phi)$ and PAS heatmap for [top] O2I [bottom] O2O.}
    \label{fig:Fig02}
\end{figure}

In Fig. \ref{fig:Fig02}, we present the variation of received power with angular misalignment in the elevation plane $(\theta)$ and in the azimuthal plane $(\phi)$, measured with respect to the line-of-sight (LOS), i.e. the line of perfect alignment. The total misalignment angle $\psi$ of the receiver from the transmitter is given by, $\psi = \cos ^{-1}{[\cos({\theta})\cos({\phi})]}$. The effect of this total misalignment on received power can vary across scenarios (O2I/ O2O), as seen from the different trends in Fig. \ref{fig:Fig01}. For a similar overall misalignment angle, O2I and O2O show different features; specifically, there are dominant clusters in the O2O case.

On the other hand, as seen from Fig. \ref{fig:Fig02}, for a given scenario, there exists a strong correlation of power angle profiles (PAPs) for similar absolute elevation angles. Overall, the PAPs reveal that misalignment can severely affect the received power. This motivated us to examine whether it is possible to develop a model for a directional power delay profile (PDP), $|h(\tau,\psi)|^2$, as a function of misalignment angle ($\psi$) and temporal index in the delay domain $(\tau)$, which remains fairly stationary within a given range of $\psi$.

The S-V model is a non-geometry-based statistical model to describe stochastic properties of delays and amplitudes of MPCs, where individual MPCs or rays can be grouped into clusters \cite{SV2}. Along the clusters and within a cluster, the amplitudes follow an exponential decay, whereas the delays of rays and clusters follow Poisson processes. Mathematically, the channel impulse response (CIR) is represented as \cite{ChandraTVT}
\begin{equation}
h(t) = \sum_{n=1}^{N_c} \sum_{m=1}^{N_{r,n}} \beta_{m,n} \exp(j \varrho_{m,n}) \delta(t-T_n-\tau_{m,n})
\label{eqy}
\end{equation} 
where $N_c$ is the number of clusters, $N_{r,n}$ is the number of rays in the $n$th cluster, and $T_n$ is the arrival time of the $n$th cluster. The magnitude, phase, and additional delay of the $m$th ray within the $n$th cluster are given by $\beta_{m,n}$, $\varrho_{m,n}$, and $\tau_{m,n}$, respectively. The inter- and intra-cluster exponential decay rates, namely $\Gamma$ and $\gamma$, define the magnitude of individual rays according to
\begin{equation}
\beta_{m,n}^2 = \beta_{1,1}^2 \exp[-(T_n-T_1)/\Gamma] \exp(-\tau_{m,n}/\gamma)
\end{equation}  
The duration of a cluster and the duration between rays within a cluster follow exponential distributions 
\begin{subequations}
\begin{gather} 
\mathrm{Pr}(T_n|T_{n-1}) = \Lambda \exp[- \Lambda (T_n -T_{n-1})] \label{eqi} \\
\mathrm{Pr}(\tau_{m,n}|\tau_{m-1,n}) = \lambda \exp[- \lambda (\tau_{m,n} -\tau_{m-1,n})] \label{eql}
\end{gather}
\end{subequations}  
with parameters $1/\Lambda$ and $1/\lambda$ representing the average duration of a cluster and the gap between two consecutive rays within a cluster, respectively.

\begin{algorithm}[!ht]
\label{algo1}
\DontPrintSemicolon
\Input{$S_c$, $\psi$}
\Output{$\textbf{h}$, $t$}

\eIf{$S_c == 1$} 
{Choose $\{\gamma, \Gamma, \lambda, \Lambda\}$ from Table \ref{o2i} for given $\psi$ \;}
{Choose $\{\gamma, \Gamma, \lambda, \Lambda\}$ from Table \ref{o2o} for given $\psi$ \;}

$T_n \gets 0$, $ix \gets 0$ \;
Initialize $h_t$ and $t_t$ with zero vectors \;

\For{$i \gets 1$ to $N_{c}$}
{
$T_n :=0$ \;
\While{$\tau_{m,n} < k \ast \gamma(i)$}
{
$t_{val} := T_n + \tau_{m,n}$ \;
Compute $h_{val}$ according to \eqref{eqy} \;
$ix := ix + 1$\;
$h_t(ix) := h_{val}$, $t_t(ix) := t_{val}$ \;
$\tau_{m,n}$ is updated according to \eqref{eql} \;
}
$T_n$ is updated according to \eqref{eqi} \;
}

Sort and reshape $h_t$ and $t_t$ to obtain $\breve{h}$ and $t$ \;
$h_s \sim \mathcal{N}(0,\sigma_{x}^{2})$ \;
$\textbf{h} := h_s \ast \breve{h}$ \;
\caption{Simulating directional CIR, $h(S_c,\psi)$}
\end{algorithm}

In this paper, we modify the basic S-V model to account for the misalignment angle as follows. First, the measured PDPs are normalized and major MPCs are identified by considering the peaks above the average of the PDP. Following this, we visually identify the clusters in the measured PDPs. We observe an average of 2 clusters for the PDPs from the O2I scenario and an average of 3 clusters in PDPs from the O2O scenario. Next, the four major S-V parameters, namely, ray arrival rate ($\lambda$), cluster arrival rate ($\Lambda$), ray decay rate ($\gamma$), and cluster decay rate ($\Gamma$)) are computed. The ray decay rate is calculated by fitting regression lines through the major MPCs and the cluster decay rate is calculated by fitting regression lines through the first peaks in each cluster. The fitting is done using the linear least squares criterion. The ray arrival rate is obtained as the mean of time separation between MPCs and the cluster arrival rate is obtained as the mean of the time separation between clusters \cite{chong2019sv}. Further, based on the results of \cite{beammisalignment}, we divide misalignment into ranges of (0$^\circ$, 10$^\circ$] and (10$^\circ$, 25$^\circ$]. The S-V parameters for a particular range are calculated as the mean of the previously obtained S-V parameters for each angular position in that range.

We summarize the modified S-V algorithm to generate directional CIR in Algorithm \ref{algo1}. We take two inputs, one considering the scenario and the second the misalignment. The S-V parameters based on the scenario and misalignment range are used as inputs to generate the PDPs. The parameters used are detailed in Table \ref{o2i} for the O2I scenario and in Table \ref{o2o} for the O2O scenario. 

\begin{table}[h!]\setlength\tabcolsep{3pt}
    \centering
    \caption{PDP simulation parameters for O2I scenario}
    \label{o2i}
    \begin{tabular}{|c|c|c|c|c|c|c|c|}
        \hline
         Misalignment   &  $N_c$ & $\lambda_{1}$ &  $\lambda_{2}$ & $\Lambda$ & $\gamma_{1}$ & $\gamma_{2}$ & $\Gamma$ \\
         angle $(\psi)$ &        & [1/ns]        & [1/ns]         & [1/ns]    & [ns]         & [ns]         & (ns)     \\
         \hline
         0\degree - 10\degree & 2 & 6.97 & 7.29 & 0.31 & 0.21 & 0.79 & 0.93\\
         \hline
         10\degree - 25\degree & 2 & 7.01 & 7.14 & 0.28 & 0.24 & 0.86 & 0.94\\
         \hline
         LOS & 2 & 5.88 & 5.88  & 0.26 & 0.21 & 0.58 & 0.45\\
        \hline
    \end{tabular}
\end{table}

\begin{table}[ht!]\setlength\tabcolsep{3pt}
    \centering
    \caption{PDP simulation parameters for O2O scenario}
    \label{o2o}
    \begin{tabular}{|c|c|c|c|c|c|c|c|c|c|}
        \hline
         Misalignment &  $N_c$ & $\lambda_{1}$ &  $\lambda_{2}$& $\lambda_{3}$ & $\Lambda$ & $\gamma_{1}$ & $\gamma_{2}$ & $\gamma_{3}$ & $\Gamma$ \\
         angle $(\psi)$ &&[1/ns]&[1/ns]&[1/ns]&[1/ns]&[ns]&[ns]&[ns]&[ns]\\
         \hline
         0\degree - 10\degree & 3 & 7.42 & 4.53 & 6.86 & 0.57 & 0.74 & 0.69 & 0.78 & 4.5\\
         \hline
         10\degree - 25\degree & 3 & 7.12 & 6.51 & 7.78 & 0.56 & 0.79 & 0.74 & 0.81 & 9.5 \\
         \hline
         LOS & 3 & 6.00 & 7.00 & 6.00 & 0.61 & 0.72 & 0.69 & 0.68 & 5.0\\
         \hline
    \end{tabular}
\end{table}

In our algorithm, we take the first cluster arrival ($T_1$) to occur at zero. We then run a loop for $N_{c}$ iterations. Thus, for each cluster, we take the first ray arrival ($T_n$) to occur at the beginning of their respective clusters. Within the ray loop, the time of arrival of each ray ($t_{val}$) is calculated and the index of that ray is incremented. We compute the value of $h_{val}$, CIR value according to \eqref{eqy} and update $\tau_{m,n}$ following the \eqref{eql}, while $T_n$ is updated as given by \eqref{eqi} for each cluster. To account for the long-term fading and shadowing, we introduce a log-normal random variable $h_s$. A MATLAB-based code implementation of the algorithm along with the measured dataset is made available in a GitHub repository \cite{NitishaGitHub}.

We obtain the CIR as the output of our algorithm, which is utilized to generate the PDPs. We compare the performance of simulated PDP, as obtained from the Algorithm-\ref{algo1}, and the observed PDPs in the following section.

\section{Channel Model Validation}
\begin{figure*}[ht]
    \centering
    \includegraphics[width=7in,clip,keepaspectratio]{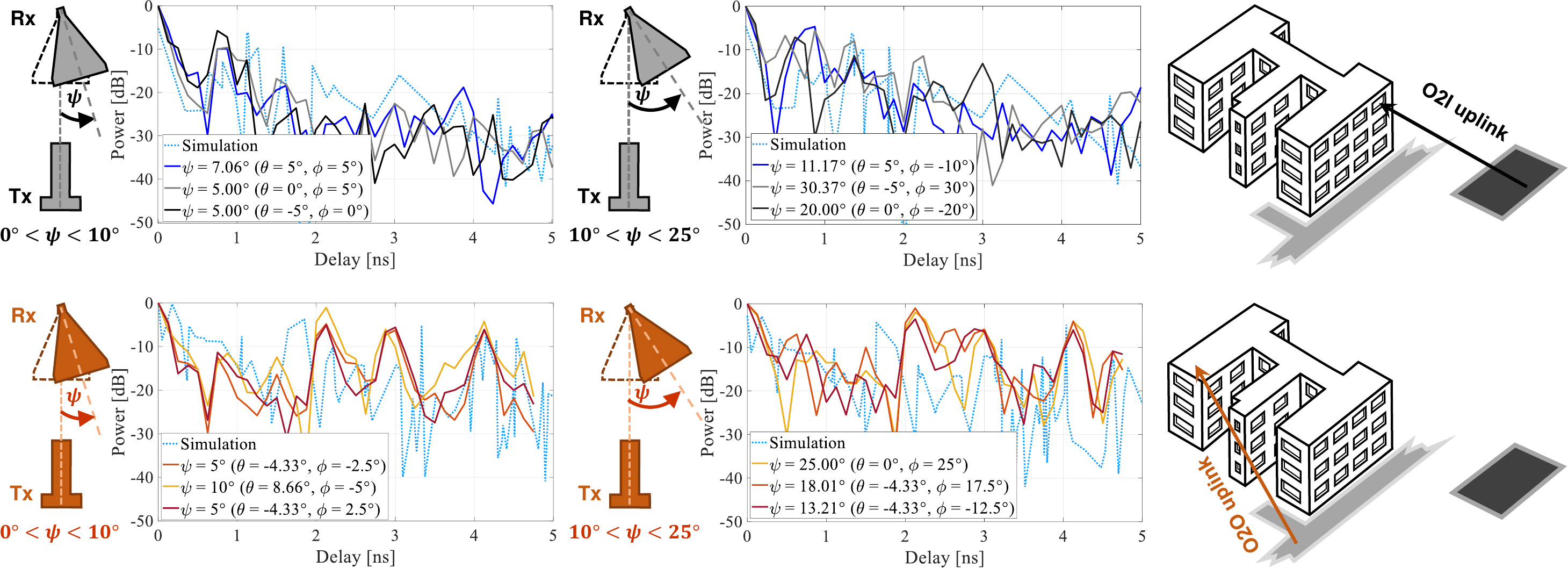}
    \caption{Comparison of normalized PDPs for O2O/ O2I uplink scenarios for two ranges of misalignment angle $(\psi)$. [left-top] O2I $(0\degree<\psi<10\degree)$, [right-top] O2I $(10\degree<\psi<25\degree)$, [left-bottom] O2O $(0\degree<\psi<10\degree)$ and [right-bottom] O2O $(10\degree<\psi<25\degree)$.}
    \label{fig:Fig03}
\end{figure*}

Validation of the algorithm proposed in the previous section is achieved through a comparison of the simulated PDPs with the measured PDPs. The goodness of fit (GoF) metrics quantifying the comparison are twofold. The first metric is the correlation $(\rho)$ \cite{correlation},
\begin{equation}
    \rho = \left|\frac{(1/N)\sum_{n = 1}^{N}|P(n)||P_s(n)|}{\sqrt{(1/N) \sum_{n = 1}^{N}|P(n)|^2 (1/N) \sum_{n = 1}^{N}|P_s(n)|^2}}\right|
\end{equation}
where $P(n)$ and $P_s(n)$ denote discretized measured and simulated PDPs of length $N$, while the second metric is two-sample Kolmogorov–Smirnov (K-S) test statistic $(\mathcal{S}_{KS})$ \cite{massey1951kolmogorov},   
\begin{equation}
    \mathcal{S}_{KS} = \max[\mathcal{F}(|P(n)|)-\mathcal{F}(|P_s(n)|)]
\end{equation}
calculated with a 5\% significance level, where $\mathcal{F}(\cdot)$ denote corresponding cumulative distribution functions. In addition to these two metrics, we also compare the RMS delay spread values. 

\begin{figure}[ht]
    \centering
    \includegraphics[width=3in,clip,keepaspectratio]{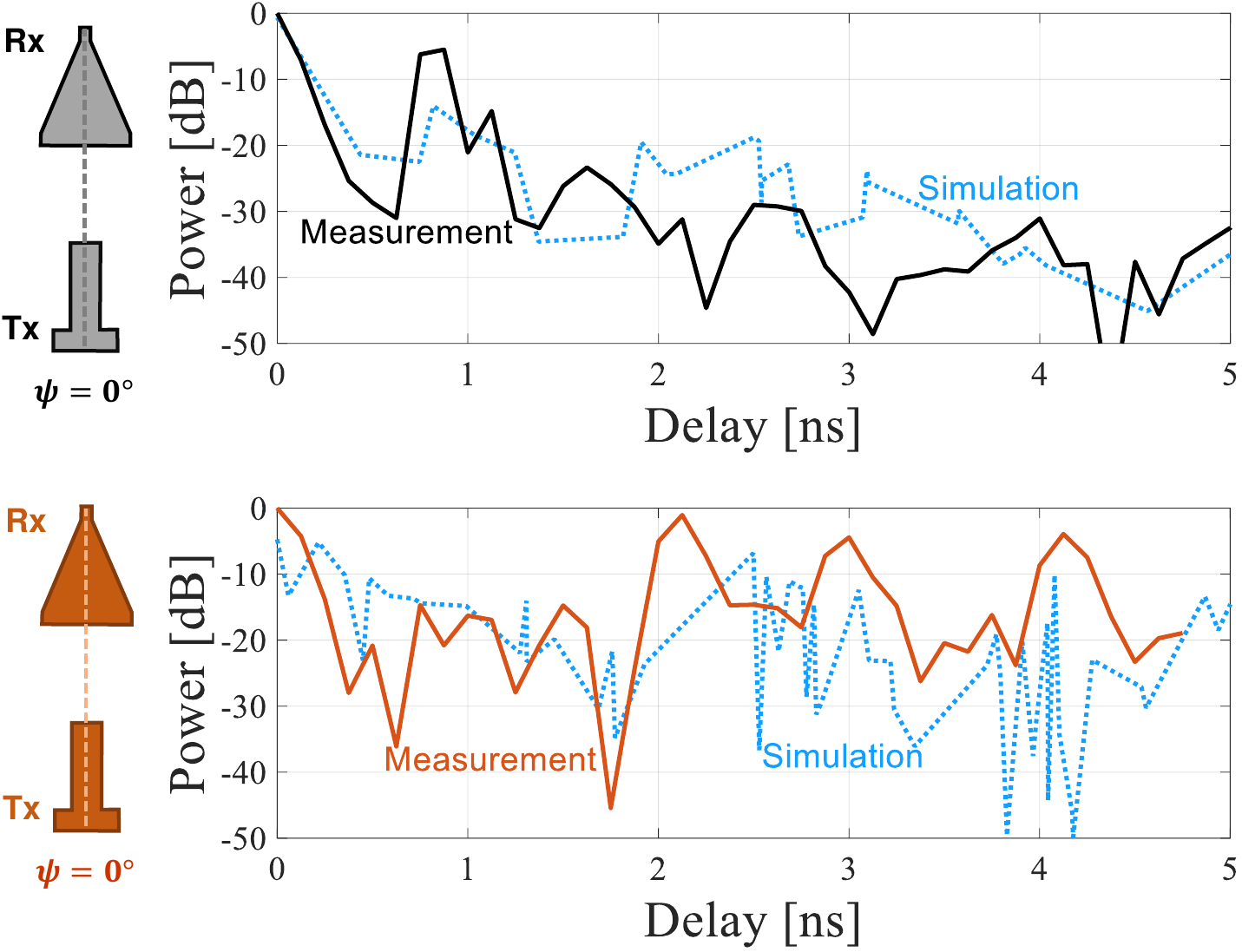}
    \caption{LOS ($\psi=0\degree$) uplink PDP comparison: [top] O2I [bottom] O2O.}
    \label{fig:Fig04}
\end{figure}

In Fig. \ref{fig:Fig03}, we depict a comparison of three sets of measured PDPs for a given range of misalignment angle with the PDP simulated for that angular range. The simulated PDP (dotted line) is able to follow the average trend slope for all three instances and can capture most of the major peaks, i.e., the multi-path components (MPCs). The claim is further demonstrated in Fig. \ref{fig:Fig04}, where we depict the LoS (no misalignment) cases and compare the measured and simulated PDPs.


Table \ref{table_3} compares the RMS delay spread of the simulated PDPs with the actual PDPs for the O2I scenario and also presents the GoF metrics. The RMS delay spread of the generated PDP lies within 6\% of the actual RMS delay spread values.
\begin{table}[ht!]\setlength\tabcolsep{3pt}
\centering
\caption{GoF and RMS delay spread comparison for O2I scenario}
\label{table_3}
\begin{tabular}{|c|c|c|c|c|c|c|}
\hline
\multicolumn{3}{|c|}{Misalignment angle}&\multicolumn{2}{|c|}{GoF}&\multicolumn{2}{|c|}{RMS delay spread [ns]}\\
\hline
$\psi$ & $\theta$ & $\phi$ & Correlation & K-S test & Measurement &  Simulation \\ \hline
 7.06\degree &  5\degree &   5\degree & 0.93  & 0.14  & 0.70  &      \\ \cline{1-6}
 5.00\degree &  0\degree &   5\degree & 0.94  & 0.21  & 0.66  & 0.86 \\ \cline{1-6}
 5.00\degree & -5\degree &   0\degree & 0.93  & 0.15  & 0.65  &      \\ \hline
11.16\degree &  5\degree & -10\degree & 0.93  & 0.28  & 0.94  &      \\ \cline{1-6}
30.37\degree & -5\degree &  30\degree & 0.93  & 0.32  & 0.88  & 0.99 \\ \cline{1-6} 
20.00\degree &  0\degree & -20\degree & 0.93  & 0.24  & 0.86  &      \\ \hline
\multicolumn{3}{|c|}{0\degree (LOS)}  & 0.93  & 0.22  & 0.53  & 0.65 \\ \hline
\end{tabular}
\end{table}



The comparison of the RMS delay spread values and the GoF metrics for the O2O scenario are given in Table \ref{table_4}. In this case, the RMS delay spread of the generated PDP lies within 4\% of the RMS delay spread values of the actual PDPs. 
\begin{table}[ht!]\setlength\tabcolsep{3pt}
\caption{GoF and RMS delay spread comparison for O2O scenario}
\label{table_4}
\begin{center}
\begin{tabular}{|c|c|c|c|c|c|c|}
\hline
\multicolumn{3}{|c|}{Misalignment angle}&\multicolumn{2}{|c|}{GoF}&\multicolumn{2}{|c|}{RMS delay spread [ns]}\\
\hline
$\psi$ & $\theta$ & $\phi$ & Correlation & K-S test & Measurement &  Simulation \\ \hline
 5.00\degree & -4.33\degree & -2.5\degree & 0.83  & 0.24  & 1.56  &      \\ \cline{1-6}
10.00\degree &  8.66\degree & -5.0\degree & 0.78  & 0.35  & 1.70  & 1.63 \\ \cline{1-6}
 5.00\degree & -4.33\degree &  2.5\degree & 0.81  & 0.29  & 1.57  &      \\ \hline
25.00\degree &     0\degree & 25.0\degree & 0.86  & 0.38  & 1.74  &      \\ \cline{1-6}
18.01\degree & -4.33\degree & 17.5\degree & 0.82  & 0.41  & 1.68  & 1.64 \\ \cline{1-6}
13.21\degree & -4.33\degree &-12.5\degree & 0.86  & 0.38  & 1.57  &      \\ \hline
\multicolumn{3}{|c|}{0\degree (LOS)} & 0.81  & 0.42  & 1.72  & 1.72 \\ \hline
\end{tabular}
\end{center}
\end{table}

\section{Conclusion}
In this paper, we examine 60 GHz fixed uplink channel PDPs at the edge of a 100m cell in O2I and O2O scenarios when Tx and Rx are not necessarily aligned with each other. Based on real measurements, we propose a modified SV model, and various statistical metrics show that the presented model can approximate PDPs in both O2I and O2O scenarios. Our model is a simple alternative to computationally extensive ray tracing models that often require precise knowledge of the environment geometry. Our work further shows that with small modifications, the traditional S-V model developed originally for characterizing ultra-wide-band channels in the 3-11 GHz band, can be used to model outdoor mmWave channels at 60 GHz with sufficient accuracy.

\bibliographystyle{IEEEtran}
\bibliography{reference}


\end{document}